\documentstyle[12pt,psfig,epsf]{article}
\newcommand{\be}{\begin{equation}}
\newcommand{\qee}{\end{equation}}
\newcommand{\bea}{\begin{eqnarray}}
\newcommand{\eea}{\end{eqnarray}}

\begin{document}

\hfill  NTZ 10/2000

\hfill  NRCPS-HE-2000-13

\vspace{24pt}
\begin{center}
{\large \bf Loop Transfer Matrix and Loop Quantum Mechanics
}

\vspace{24pt}
{\sl G.K.Savvidy}

Institut f\"{u}r Theoretische Physik, Universit\"{a}t Leipzig, \\
Augustusplatz 10, D-04109 Leipzig, Germany\\

and\\

National Research Center Demokritos,\\
Ag. Paraskevi, GR-15310 Athens, Greece 

\end{center}
\vspace{12pt}

\centerline{{\bf Abstract}}

\vspace{23pt}
\noindent

The gonihedric model of random surfaces on a 3d Euclidean lattice 
has equivalent representation in terms of transfer matrix 
$K(Q_{i},Q_{f})$ which describes the propagation of loops $Q$.
We extend the previous construction of loop transfer matrix to
the case of nonzero self-intersection coupling constant $\kappa$. 
We introduce loop generalization of Fourier transformation
which allows to diagonalize  transfer matrices  depending 
on symmetric difference of loops and express  
all eigenvalues of $3d$ loop transfer matrix through 
the correlation functions of the 
corresponding 2d statistical system. The 
loop Fourier transformation allows to carry out analogy with
quantum mechanics of point particles, to introduce conjugate loop momentum P 
and to define loop quantum mechanics. 

We also consider 
transfer matrix on $4d$ lattice which describes propagation of 
memebranes. This transfer matrix can also be diagonalized by using 
generalized Fourier transformation, 
and all its eigenvalues are equal to the correlation
functions of the corresponding $3d$ statistical system. Particularly 
the free energy of the $4d$ membrane system is equal to the free energy
of $3d$ gonihedric system of loops and is equal to the free energy of 2d Ising 
model.


\newpage

\pagestyle{plain}
\section{Introduction}

Various  models of  random surfaces built out of 
triangles embedded into continuous space $R^d$ 
and surfaces built out of  plaquettes  
embedded into Euclidean  lattice $Z^d$ have been considered in the literature 
\cite{weingarten,gross,karowski}.
These models  are based on area action and suffer the problem 
of non-scaling behaviour of the string tension and the dominance of 
branched polymers \cite{ambjorn}.   
Several studies have analyzed the physical effects produced 
by rigidity of the surface introduced by adding 
dimension-less extrinsic curvature term to the area action \cite{helfrich}
$$
T\cdot S(area) + \frac{1}{\alpha} F(extrinsic~curvature),
$$    
where T is a string tension, $\alpha$ is dimension-less coupling constant
and $F(ext~curv) \propto 1$ is dimension-less functional.
Comprehensive review of work in this area up to 1997 can be 
found in \cite{ambjorn}. 

In \cite{amb,sav} authors suggested  so-called gonihedric model 
of random surfaces which is also based on the 
concept of extrinsic curvature, but it differs in two essential points 
from the models considered in the previous 
studies. First it claims  
that only extrinsic curvature term should be considered 
as the fundamental action of the theory 
$$
S =m \cdot A(extrinsic~curvature),
$$
here $m$ has dimension of mass and there is no area term in the action. 
Secondly it is required that the dependence on the
extrinsic curvature should be such that the action will   
have dimension of length  
$
A(extrinsic~curvature) ~~ \propto ~~length.
$ 
This means that the action should measure the surfaces in terms of length
and it should be proportional to the linear size of the surface,
as it was for the path integral. The last property will guarantee that  
in the limit when the surface degenerates into a single world line   
the functional integral over surfaces will naturally transform into the   
Feynman path integral for point-like relativistic particle (see Figure 1,2) 
\footnote{This is in contrast with the previous 
proposals when the extrinsic curvature 
term is a dimension-less functional $F(extrinsic~curvature) \propto 1$
and can not provide this property.}
$$
m ~A_{xy}(extrinsic~curvature)~~ \rightarrow ~~ m \int^{y}_{x} dl .
$$
As it was demonstrated in \cite{sav} 
quantum fluctuations  generate the area term in the effective 
action
$$
S_{eff} = m~A(extrinsic~curvature)~~ + ~~T_{eff} ~ S(area)~~ +~~ ...,
$$
that is nonzero string tension  $T_{eff}$.
Therefore at the tree level the theory describes free quarks with string
tension equal to zero, but now quantum fluctuations generate nonzero string 
tension and, as a result, quark confinement \cite{sav}. The theory 
may consistently describe asymptotic freedom and confinement 
as it is expected to be the case in QCD. 

Our aim is to 
find out what the continuum limit of this regularized theory of 
random surfaces is, and if indeed the continuum limit 
corresponds to nontrivial string theory, does it describe QCD 
phenomena. This is a complicated problem and one of the possible
ways to handle the problem is to go further and to formulate the theory 
also on a lattice. An equivalent representation of the model 
on Euclidean lattice $Z^{d}$ has been formulated in \cite{weg}. This
representation depends on dimension $d$ of the embedding space. In 
three dimensions it is a spin model in which the interaction 
between spins is organized in a very specific way (see equation 
(\ref{hamil}) ).In high dimensions
it is a gauge spin system which contains one, two and  three plaquette
interaction terms.

In this article I shall consider the above model of random 
surfaces embedded into 3d Euclidean lattice $Z^3$. The reason to focus 
on that particular case is motivated by the fact that one can 
geometrically construct the corresponding transfer matrix  \cite{sav2} 
and find its exact spectrum \cite{transferma}. 
The spectrum of the transfer matrix which depends 
only on symmetric difference of initial and final loops
has been evaluated exactly in terms 
of correlation functions of the 2d Ising model in \cite{transferma}.
This transfer matrix has the form \cite{sav2}
$$
K_{\kappa =0}(Q_{1},Q_{2}) = 
exp \{-\beta ~[ k(Q_{1}) + 
2 l(Q_{1} \bigtriangleup Q_{2}) + k(Q_{2})]~ \},
$$
where $Q_{1}$ and $Q_{2}$ are closed polygon-loops on a two-dimensional 
lattice, $k(Q)$ is the curvature and $l(Q)$ is
the length of the polygon-loop $Q$ \footnote{We shall use the 
word "loop" for the "polygon-loop".}. 
This transfer matrix describes  the propagation 
of the initial loop $Q_{1}$ to the final loop $Q_{2}$. 

\begin{figure}
\centerline{\hbox{\psfig{figure=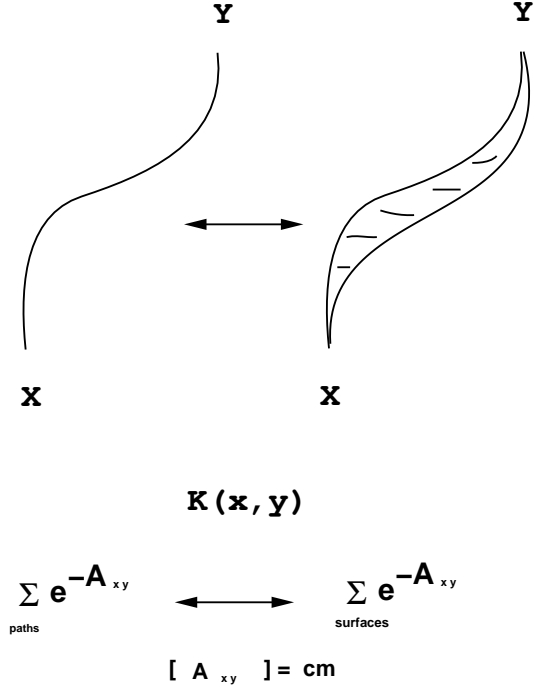,width=7cm}}}
\caption[fig]{ It is required that the action $A_{xy}$ should  
measure the surfaces in terms of length, as it was for the path integral.
When a surface degenerates into a single world line we have 
natural transition to a path integral . 
}
\label{fig}
\end{figure}

Our aim in this article is to extend these results in several 
directions. First to construct the transfer matrix for 
more general case of nonzero self-intersection coupling constant
$\kappa$
$$
K(Q_{1},Q_{2})~~=~~K_{\kappa =0}(Q_{1},Q_{2})~\cdot
exp \{-4\kappa \beta ~[ k_{int}(Q_1) + 2 l(Q_{1}  
\stackrel{\leftrightarrow}{\cap}  Q_{2}) +  k_{int}(Q_2) ]~\},
$$
where $k_{int}(Q)$ is number of self-intersection vertices of the loop $Q$.
For $\kappa =0$ this expression  coincides with the previous one.
In the limit $\kappa \rightarrow \infty,$ one can see
that on every time slice there will be no loops  with 
self-intersections and we  have propagation of nonsingular 
self-avoiding oriented loops. 

Secondly, I will demonstrate that diagonalization of any transfer
matrices $K(Q_{1} \bigtriangleup  Q_{2})$, which depend only 
on symmetric difference of loops $Q_{1} \bigtriangleup  Q_{2}$ can be 
performed by using generalization of Fourier transformation in
loop space. This transformation will be defined by using the loop 
eigenfunctions 
$$
\Psi_{P}(Q) = e^{i\pi s(P \cap Q) } 
$$
which are numbered by  the {\it loop momentum} $P$ and $s(Q)$ is the 
area of the region with boundary loop $Q$.
They are in a pure analogy with plane waves in quantum mechanics 
$\psi_{p}(x) = e^{ipx}$. We shall prove that any transfer
matrix $K(Q_{1} \bigtriangleup  Q_{2})$ can be 
diagonalized by using this loop Fourier transformation  
$$
K(P_1 , P_2) = \sum_{\{Q_1 , Q_2\}} 
K(  Q_{1} \bigtriangleup  Q_{2})~e^{i\pi s(P_1 \cap Q_1)  -  i\pi s(P_2 \cap Q_2) } 
= \Lambda_{P_1}~ \delta (P_1 , P_2),
$$
here $\Lambda_{P}$ are eigenvalues depending on loop-momentum $P$.
Because the transfer matrix plays the role of the evolution operator, one 
can define a corresponding Hamiltonian operator as 
$
H(P) = - ln \Lambda(P).
$
At this point one can talk about {\it loop quantum mechanics } 
which should be defined by using conjugate loop operators 
$\hat{Q}$ and $\hat{P}$ and the Hamiltonian  $H(\hat{P},\hat{Q})$.

I shall also consider the 
model of  random three-dimensional   
manifolds which are build by gluing together tetrahedra in continuous 
space $R^{4}$ and  out of 3d cubes in 4d  Euclidean 
lattice $Z^4$. This system has very  
close nature with the gonihedric  model  because 
in both cases the action is proportional to the 
linear size of the manifold \cite{sav2}. 
In \cite{sav2} the authors where able to construct 
a transfer matrix for  this theory as well. This 
transfer matrix describes the propagation 
of two-dimensional membrane $M$ and our main result is that 
both theories can be solved exactly by using  generalization 
of the Fourier transformation in loop and in membrane spaces.
They represent two nontrivial theories in  
three and four dimensions respectively which can be 
solved exactly. 

In the next sections I shall present background material, the
definition of the system of random surfaces in continuous space 
as well as on Euclidean lattice and describe its basic properties. 
In the third section 
the construction of the transfer matrix for the loops will be extended to
the case of nonzero self-intersection coupling constant $\kappa$. 
In the fourth main section we shall introduce 
the generalization of Fourier transformation in the 
loop space and demonstrate that loop Fourier transformation
allows to diagonalize all transfer matrices which depend 
only on symmetric difference of loops. 
The net result is that all eigenvalues $\Lambda_{P}$ of 3d transfer matrix 
are exactly equal to the loop correlation functions of the corresponding
2d system
\be
({\Lambda_{P_{\vec{r_1},...\vec{r_n}}}  
\over \Lambda_{\emptyset} } )^{3d} ~ = ~ <\sigma_{\vec{r_1}}\cdot 
\cdot \cdot \sigma_{\vec{r_n}}>_{2d}.
\qee 
In this formula the loop momentum $P$ is defined by the 
vectors $\vec{r_1},...\vec{r_n}$ (see Figure 8). In its form the last 
relation is very close with the ones discussed in \cite{witten,maldacena1,maldacena2}.

\begin{figure}
\centerline{\hbox{\psfig{figure=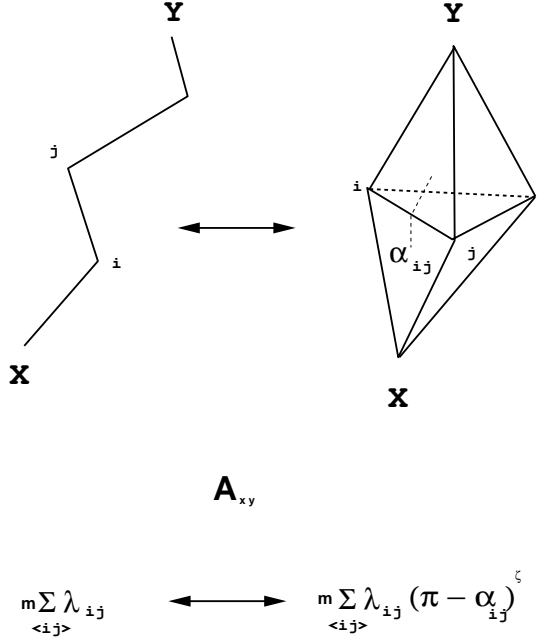,width=7cm}}}
\caption[fig1]{ The random path and random surface with boundary  
points $X$ and $Y$. In both cases the action is proportional to the linear size 
of the geometrical fluctuation \cite{sav}.}
\label{fig1}
\end{figure}

In the last section I shall consider the transfer 
matrix for membranes. This matrix also can be diagonalized by Fourier 
transformation and all its egenvalues are equal to the correlation
functions of the corresponding 3d system. In particular we shall see 
that free energy of the membrane system is equal to the free energy
of gonihedric system of loops and is equal to the free energy of 2d Ising 
model.

\section{Random Surfaces and Path Integral}

{\it Surfaces in Continuous  Space.} Gonihedric string has been 
defined as a model of random surfaces with 
the action which is proportional to the linear size of the surface 
\cite{amb,sav}

\be
m ~ \sum_{<ij>} \lambda_{ij}
\cdot \vert \pi - \alpha_{ij} \vert^{\zeta} ~~~~~~~~\zeta \leq (d-2)/d,  \label{action}
\qee
where $\lambda_{ij}$ is the length of the edge $<ij>$ of the 
triangulated surface $M$ and $\alpha_{ij}$ is the dihedral angle 
between two neighboring triangles of $M$ sharing a common edge $<ij>$ (see Figure 1,2)
\footnote{The angular factor defines the rigidity of the random surfaces 
and for $\zeta \leq (d-2)/d$ it increases sufficiently fast 
near angles $\alpha = \pi$ to suppress transverse fluctuations \cite{sav,durhuus}.}.
The action was defined for the self-intersecting surfaces 
as well \cite{sav}, it 
accounts self-intersections of different orders by ascribing 
weights to self-intersections. These weights are proportional 
to the length of the 
intersection multiplied by the angular factor which is equal to the sum of 
all angular factors corresponding to dihedral angles in the 
intersection \cite{sav} (see Figure 3)
\be
\kappa~m ~\sum_{<ij>} \lambda_{ij}
\cdot (~\vert \pi - \alpha^{1}_{ij} \vert^{\zeta}+...+
\vert \pi - \alpha^{r}_{ij} \vert^{\zeta}~ ).  \label{intersec}
\qee
The number of terms inside parentheses depends on the order of the intersection
$r$, where $r$ is the number of triangles sharing the edge $<ij>$. Coupling 
constant $\kappa$ is called self-intersection coupling constant \cite{sav}.
Thus the total action is equal to 
\be
A(M) = m~\sum_{<ij>} \lambda_{ij}
\cdot \vert \pi - \alpha_{ij} \vert^{\zeta}~~ + ~m~\kappa~\sum_{<ij>} \lambda_{ij}
\cdot (~\vert \pi - \alpha^{1}_{ij} \vert^{\zeta}+...
+\vert \pi - \alpha^{r}_{ij} \vert^{\zeta}~) \label{totaction}.
\qee
Both terms in the action have the same dimension and the 
same geometrical nature, the action (\ref{action})  measures 
two-dimensional surfaces in terms of length, self-intersections
are also measured in terms of length (\ref{intersec}).
If $\kappa$ is small then self-intersections are permitted, 
if $\kappa$ increases then there is 
strong repulsion along the curves of self-intersections, and the surfaces tend 
to be  self-avoiding \cite{sav}.

The model is a natural extension of the  Feynman
path integral for point-like relativistic particle (see Figure 1,2). 
This can be seen from 
(\ref{action}) in the limit when the surface degenerates into a single 
world line  \cite{sav}\footnote{This property of the gonihedric action 
guarantees that spike instability 
does not appear here because the action is proportional 
to the total length of the 
spikes and suppresses the corresponding fluctuations \cite{amb,sav}.}

$$~~~~~~~~~~~~\sum_{<ij>} \lambda_{ij} \cdot 
\vert \pi - \alpha \vert^{\varsigma}~~ 
\longleftrightarrow ~~ \sum_{<ij>} \lambda_{ij}.  \label{degen}$$

\begin{figure}
\centerline{\hbox{\psfig{figure=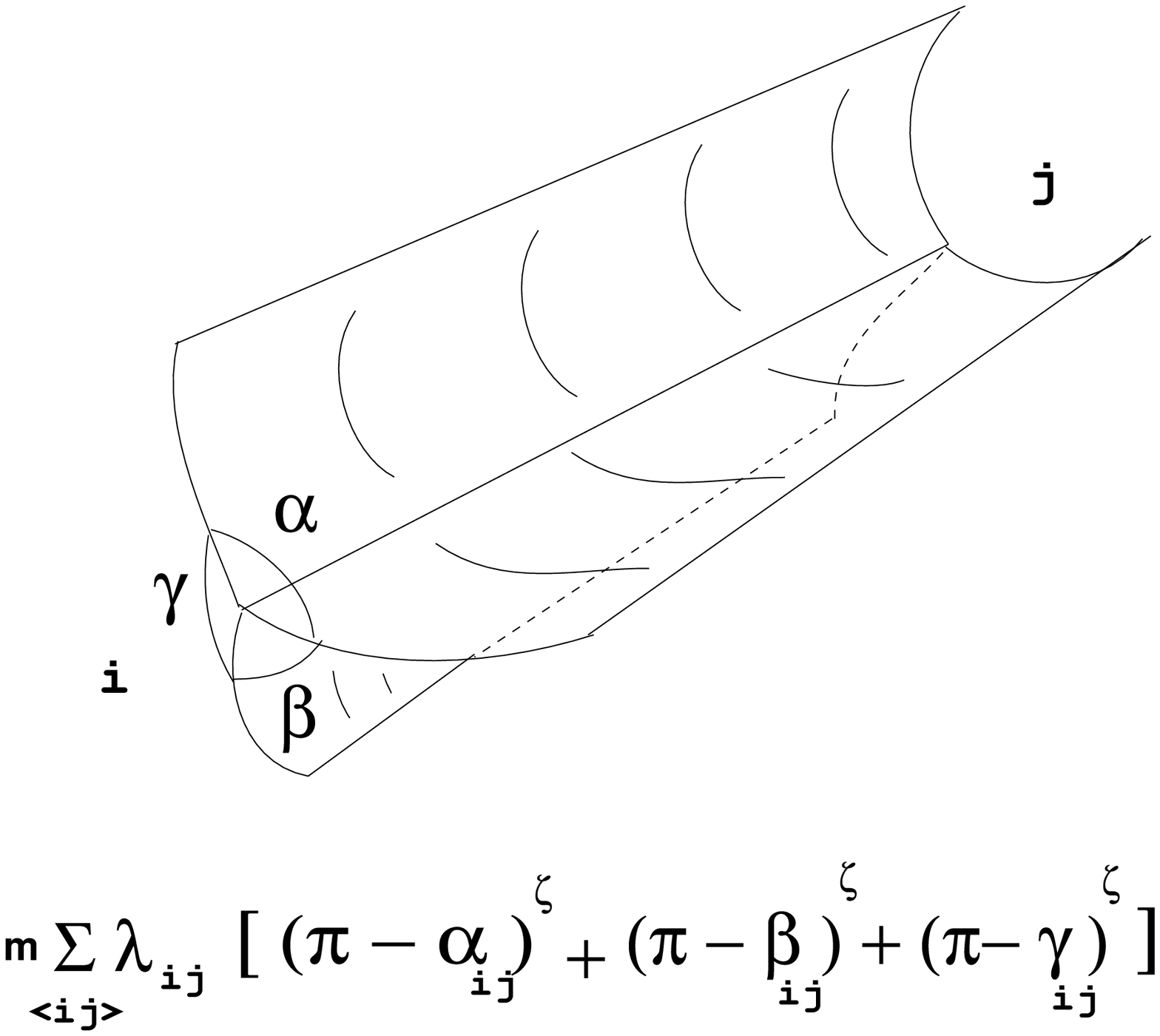,width=7cm}}}
\caption[fig2]{ The figure depicted the self-intersection of three planes
and the associated action. In the given case $\alpha + \beta +\gamma =
2\pi$.}
\label{fig2}
\end{figure}
At the classical level the string tension is equal to zero and 
quarks viewed as open 
ends of the surface are propagating freely without interaction  
$\sigma_{classical}=0$.
This is because the  action (\ref{action}) 
is equal to the perimeter of the flat Wilson loop 
$
A(M)  \rightarrow  m(R+T).
$
As it was demonstrated in \cite{sav}, quantum fluctuations generate  
nonzero string tension 
$
\sigma_{quantum}= \frac{d}{a^2}~(1 - ln \frac{d}{\beta}) ,
$
where $d$ is the dimension of the spacetime, $\beta$ is the coupling constant,
$a$ is a scaling parameter and $\varsigma = \frac{d-2}{d}$
in (\ref{action}).
In the scaling limit $\beta \rightarrow \beta_{c}=d/e$ the string tension 
has a finite limit.

We shall also consider similar  model of random membranes with the 
action which is proportional to the linear size of the 
corresponding worldvolume.
This action for three-dimensional manifold has been defined as 
\be
A(N) = m ~\sum_{<i,j>} \lambda_{ij} \cdot \vert 2 \pi - \omega_{ij} \vert^{\zeta} 
\qee
where $\lambda_{ij}$ is the length of the edge $<ij>$ of 
three-dimensional manifold $N$ which is constructed 
by gluing together three-dimensional tetrahedra 
through their triangular faces and $2\pi - \omega_{ij}$
is the deficit angle on the edge $<ij>$, 
here $\omega_{ij}$ is the sum of the 
dihedral angles between triangular faces of tetrahedra
which have the common edge $<ij>$. 

{\it Equivalent spin system on a lattice \cite{weg}. }
In addition to the formulation of the theory in the continuum space 
the system allows an equivalent representation on Euclidean lattices 
where a surface is associated with a collection of plaquettes 
\cite{weg}. 
The corresponding  statistical weights are proportional to the 
total number of non-flat edges $n_2(M)$ of the surface $M$ 
as it is defined in (\ref{action}) 
\cite{weg}. The edges are non-flat when 
the dihedral angle between plaquettes is equal to $\pi/2$. The surfaces may 
have self-intersections in the form of four plaquettes intersecting on a link. 
The weight associated with self-intersections is proportional to $4 \kappa n_4$ 
where $n_4(M)$ is the number of edges with four intersecting
plaquettes, and $\kappa$ is the self-intersection coupling constant (\ref{intersec}) .
The edges of the surface with 
self-intersections comprise the {\it singular part} of the surface. 
The edges of the surface where only two plaquettes are intersecting 
comprise the {\it regular part} of the surface. 

The partition function can be 
represented as a sum over two-dimensional surfaces $M$ of the 
type described above, embedded into a three-dimensional lattice:
\be
Z(\beta) = \sum_{\{surfaces~M\}} e^{-2 \beta~A(M)},  \label{partfan}
\qee
where $A(M)$ is the energy of the surface $M$ constructed from plaquettes
\be
A_{gonihedric}(M) = n_{2}(M) + 4\kappa n_{4}(M).   \label{energyk}
\qee

To study statistical and scaling properties 
of the system one can directly simulate surfaces by gluing 
together plaquettes following the rules described 
above, or to use the duality between random surfaces and spin systems on 
Euclidean lattice. This duality allows to  study  equivalent spin 
system with specially adjusted interaction between spins \cite{weg}.
In three dimensions the corresponding Hamiltonian is equal to 
\cite{weg}
\be
H_{gonihedric}^{3d}=- 2\kappa \sum_{\vec{r},\vec{\alpha}} \sigma_{\vec{r}}
\sigma_{\vec{r}+\vec{\alpha}}
+ \frac{\kappa}{2} \sum_{\vec{r},\vec{\alpha},\vec{\beta}} \sigma_{\vec{r}}
\sigma_{\vec{r}+\vec{\alpha} +\vec{\beta}}
-  \frac{1-\kappa}{2} \sum_{\vec{r},\vec{\alpha},\vec{\beta}} \sigma_{\vec{r}}
\sigma_{\vec{r}+\vec{\alpha}} \sigma_{\vec{r}+\vec{\alpha}+\vec{\beta}}
\sigma_{\vec{r}+\vec{\beta}}, \label{hamil}
\qee
where $\vec r$ is a three-dimensional vector on the lattice 
$Z^{3}$, the  components of which are integer and $\vec \alpha$, $\vec 
\beta$ are unit vectors parallel to the axes. The interface energy 
of this lattice spin system coincides with the linear action 
(\ref{totaction}), (\ref{energyk}).

The form of the Hamiltonian $H(\kappa)$
and the symmetry of the system essentially depend on $\kappa$: 
when $\kappa \neq 0$  one can flip the spins on arbitrary parallel layers 
and thus the degeneracy of the ground state is equal to $3\cdot 2^{N}$, 
where $N^3$ is the size of the lattice. When $\kappa=0$ the system
has even higher symmetry, {\it all states, including the ground
state are exponentially degenerate}. This degeneracy of the ground
state is equal to $2^{3N}$ \cite{sav2}. This is  because now 
{\it one can flip the spins on arbitrary layers, even on
intersecting ones}. The corresponding 
Hamiltonian  contains only exotic four-spin interaction term. 

The understanding of critical behaviour of this system comes 
from the analysis of the transfer matrix \cite{sav2} which describes the 
propagation of the closed loops $Q$ in time direction with an 
amplitude which is proportional to the sum of the length $l(Q)$ of the loop  
and of its total curvature $k(Q)$. Because the
amplitude essentially depends on length, as it takes 
place  in 2d Ising model, this result
directs to the idea 
that the system in 3d will show the phase transition which should 
be of the same nature as it is in 2d Ising system 
and that the transfer matrix describes the propagation of an 
almost free string of Ising model \cite{sav2}.
Direct numerical Monte-Carlo simulations \cite{bathas} 
and analytical   
solution of different approximations to the full transfer matrix 
confirm strong connection with the 2d Ising model \cite{transferma}.                        

Below I shall remind the derivation of 
the corresponding transfer matrix for the case $\kappa=0$, 
which is based on geometrical 
theorem proven in \cite{schneider} and then I shall derive the transfer 
matrix for more general case $\kappa \neq 0$. This will allow 
to study physical picture of string 
propagation which follows from the transfer matrix approach.

\section{Loop Transfer Matrix}

{\it Geometrical Theorem \cite{schneider}.}
In this section I shall review the construction of the transfer 
matrix for the system (\ref{hamil}) with $\kappa =0$ \cite{sav2}
\be
H_{gonihedric}^{3d}(\kappa=0)=
-  \frac{1}{2} \sum_{\vec{r},\vec{\alpha},\vec{\beta}} \sigma_{\vec{r}}
\sigma_{\vec{r}+\vec{\alpha}} \sigma_{\vec{r}+\vec{\alpha}+\vec{\beta}}
\sigma_{\vec{r}+\vec{\beta}}, \label{nullhamil}
\qee
and then extend the result to nonzero case $\kappa \neq 0$.

In order to find transfer matrix for this system we have to use geometrical
theorem proven in \cite{schneider}.
The geometrical theorem  provides  an equivalent 
representation of the action $A(M)$ in terms of total curvature
$k(E)$ of the polygons which appear in the intersection of the
two-dimensional plane $E$ with the given two-dimensional 
surface $M$ (see Figure 4,5).
This curvature $k(E)$ should then be integrated over all planes $E$ 
intersecting the surface $M$. 

\begin{figure}
\centerline{\hbox{\psfig{figure=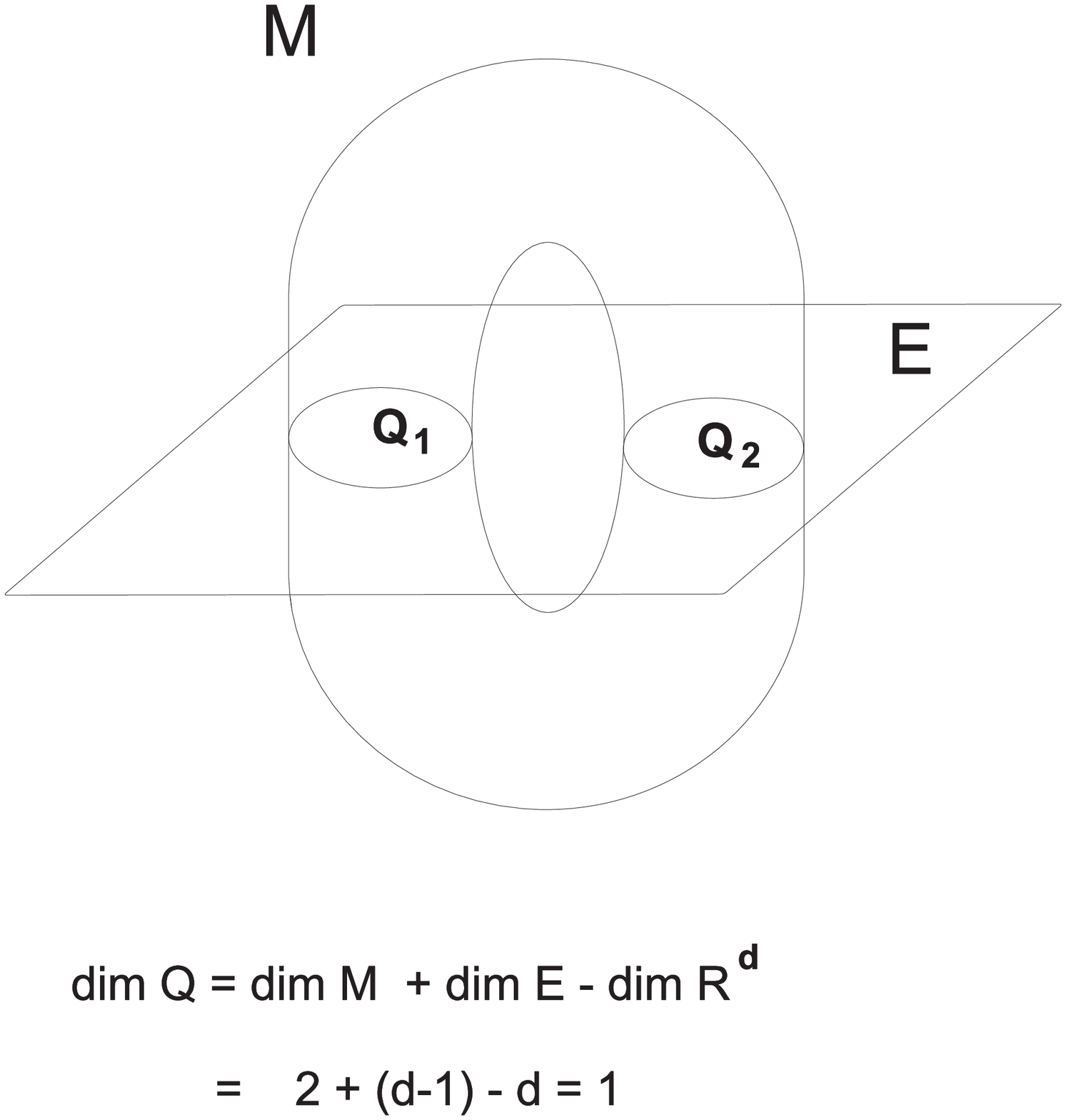,width=7cm}}}
\caption[fig]{The geometrical theorem provides  an equivalent representation 
of the action $A(M)$ in terms of total curvature
$k(E)$ of the curves $Q_{1}(E),...,Q_{k}(E)$ which appear in the intersection of the
(d-1)-dimensional plane $E$ with the given two-dimensional surface $M$.
The curvature $k(E)$ should then be integrated over all planes $E$ 
intersecting the surface $M$.  
}
\label{fig}
\end{figure}

The intersection of the two-dimensional
plane $E$ with the polyhedral surface $M$ is the union of the polygons
(see Figure 5) $$Q_{1}(E) , ...,Q_{k}(E) $$
and the absolute total curvature $k(E)$ of these  polygons is equal to 
\be
 k(E) = \sum^{k}_{i=1} k(Q_{i}) =
\sum_{<i,j>} \vert \pi-\alpha^{E}_{ij} \vert , \label{pol}
\qee
where $\alpha^{E}_{ij}$ are the angles of these polygons. They are
defined as the angles in the intersection of the two-dimensional
plane $E$ with the edge $<ij>$ \cite{schneider} (see Figure 5).
The meaning of (\ref{pol}) is
that it measures the total revolution of the
tangent vectors to polygons  $Q_{1}(E) , ...,Q_{k}(E)$.
Integrating the total curvature $k(E)$ (\ref{pol}) over all
intersecting planes $E$ we shall get the action $A(M)$ \cite{schneider}
\be
A(M) = \frac{1}{2\pi}\int_{\{ E \} }k(E)dE  \label{curv}.
\qee

\begin{figure}
\centerline{\hbox{\psfig{figure=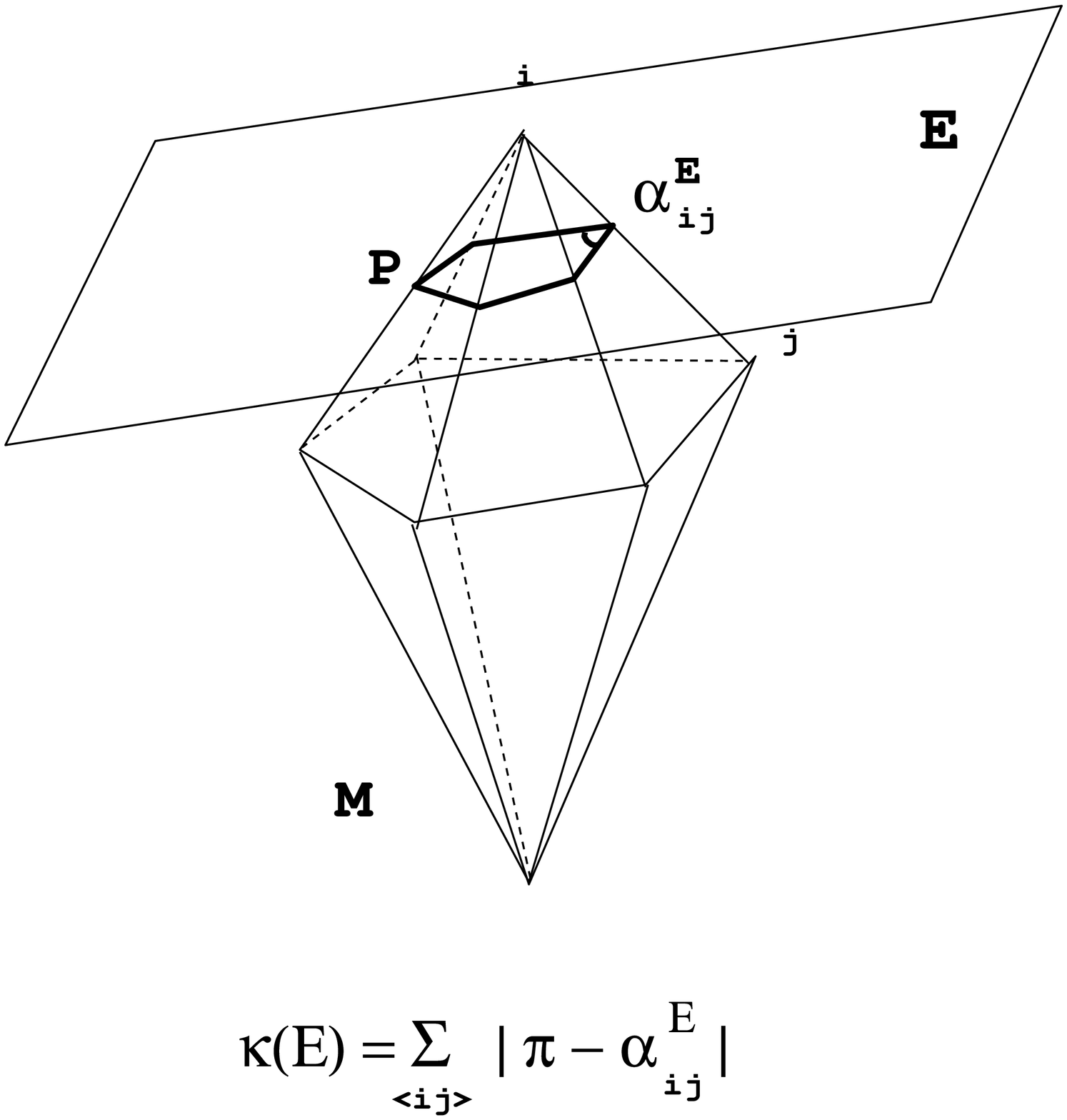,width=7cm}}}
\caption[fig]{ Intersection of the polyhedral surface $M$
by the plane $E$. The image of the surface on the plane is a 
polygon $P$ and its curvature is equal to $k(E)$.
}
\label{fig}
\end{figure}

{\it Geometrical Theorem on a lattice \cite{sav2}.}
One can find the same representation (\ref{curv}) for the action
$A(M)$ on a cubic lattice $Z^{3}$ introducing a set of planes 
$\{ E_{x} \}, \{ E_{y} \}, \{ E_{z} \} $ perpendicular to $x, y,z $ axis
on the dual lattice $Z^{3}_{d}$.
These planes will intersect a given surface $M$ and
on each of these planes we shall have an image of the  surface
$M$. Every such image is represented as a collection of
closed polygons $ Q(E)$ appearing in the intersection of
the plane with surface $M$ (see Figure 6).
The energy of the surface $M$ is equal now to the
sum of the total curvature $k(E)$ of all these polygons
on  different planes
\be
A(M) = \sum_{\{ E_{x},E_{y},E_{z}\} }
k( E ). \label{latcurv}
\qee
The total curvature  $k(E)$ is simply the total number of polygon right angles. 
In the case when the self-intersection coupling constant $\kappa=0$
is equal to zero   we must compute the total curvature $k(E)$ of
these polygons ignoring angles at the self-intersection points!

With (\ref{latcurv}) the partition function of the system can 
be written in the form
\be
Z(\beta) = \sum_{\{M\}}
exp\{-2 \beta \sum_{\{E\}} k(E)\}    \label{partfun} 
\qee
where the sum in the exponent can be represented as a product
\be
exp\{-2 \beta \sum_{\{E\}} k(E)\} =
\prod_{\{E\}}  e^{-2 \beta k(E)} =
\prod_{\{E_{z}\}}  e^{-2 \beta k(E_{z})}
\prod_{\{E_{y}\}}  e^{-2 \beta k(E_{y})}\prod_{\{E_{x}\}}
e^{-2 \beta k(E_{x})}. \label{product}
\qee
The goal is to express the energy functional (\ref{latcurv}) and the
product (\ref{product}) in terms of quantities defined only on
two-dimensional planes in one fixed direction, let's say through $\{E_{z}\}$.

{\it Transfer Matrix  for $\kappa=0$ \cite{sav2}. }
The question is: what kind
of information do we need to know on these planes  $\{E_{z}\}$
to recover the values of the total curvature $k(E_{x})$
and $k(E_{y})$ on the planes  $\{ E_{y}\}$ and $\{E_{x}\}$
for the given surface $M$? 

Let us consider the sequence of two planes
$E_{z}^{i}$ and  $E_{z}^{i+1}$ and denote by $Q_{i}$
the polygon-image of the surface $M$ on the plane  $E_{z}^{i}$
and by $Q_{i+1}$ the polygon-image of $M$ on the plane
$E_{z}^{i+1}$. 
The  contribution to the curvature $k(E_{x})+k(E_{y})$
of the polygons which are on the perpendicular planes between
$E^{i}_{z}$ and $E^{i+1}_{z}$  is equal to the length of the polygons
$Q_{i}$ and $Q_{i+1}$ without length of the common bonds \cite{sav2}
\be
l(Q_{i})+l(Q_{i+1})-2 \cdot l(Q_{i} \cap Q_{i+1} ) 
= l(Q_{i} \bigtriangleup  Q_{i+1}) , \label{common}
\qee
where the polygon-loop~ $Q_{1} \bigtriangleup  Q_{2}~ \equiv~ Q_{1} \cup  
Q_{2} ~\backslash~ Q_{1} \cap  Q_{2} $~ is a union of links 
$Q_{1} \cup Q_{2}$ without common links $Q_{1} \cap  Q_{2}$. 
Therefore (see formula (\ref{latcurv}))
\be
A(M) = \sum_{\{ E\}} k(E) = \sum_{\{ E_{z} \}}
k(Q_{i}) + l(Q_{i} \bigtriangleup  Q_{i+1}). \label{commonact}
\qee
The partition function (\ref{partfun}) can be now represented in the form
\be
Z(\beta) = 
\sum_{\{Q_{1},Q_{2},...,Q_{N}\}}~ K_{\beta}(Q_{1},Q_{2})\cdots 
K_{\beta}(Q_{N},Q_{1})  = tr K^{N}_{\beta}\label{trace},
\qee
where $K_{\beta}(Q_{1},Q_{2})$ is the transfer matrix of  size 
$\gamma \times \gamma $, defined as 
\be
K^{gonihedric}_{\kappa =0}(Q_{1},Q_{2}) = 
exp \{-\beta ~[k(Q_{1}) + 
2 l(Q_{1} \bigtriangleup  Q_{2}) + k(Q_{2})]~ \},  \label{tranmat}
\qee
where $Q_{1}$ and $Q_{2}$ are closed polygon-loops on a two-dimensional 
lattice 
$T^{2}$ of size $N \times N$ and $\gamma = 2^{N^2}$ is the total number of polygon-loops 
on a lattice $T^{2}$. 
The transfer matrix (\ref{tranmat}) can be viewed as describing  the propagation 
of the polygon-loop $Q_{1}$ at time $\tau$ to another polygon-loop $Q_{2}$ 
at the time $\tau +1$ (see Figure 7)\footnote{Layer-to-layer transfer matrices for 
three-dimensional statistical systems
have been considered in the literature 
\cite{polyakov,zamolodchikov}. 
Using Yang-Baxter and Tetrahedron equations 
one can compute the spectrum of the transfer matrix in a number of 
cases \cite{zamolodchikov}. In the given case the transfer matrix 
has geometrical interpretation which helps to compute the 
spectrum.}.

\begin{figure}
\centerline{\hbox{\psfig{figure=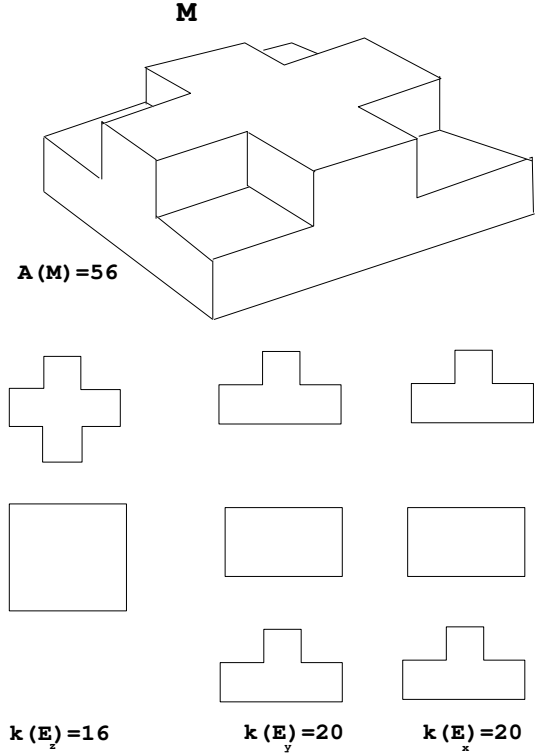,width=7cm}}}
\caption[fig]{ The set of planes 
$\{ E_{x} \}, \{ E_{y} \}, \{ E_{z} \} $ perpendicular to $x, y,z $ axis
intersect a given surface $M$ in the middle of the links.  
On each of these planes we shall have an image of the  surface
$M$. Every such image is represented by closed polygon-loops 
$ P(E)$ appearing in the intersection of
the plane with surface $M$.
The energy of the surface $M$ is equal to the
sum of the total curvature $k(E)$ on all these planes as it is 
given by the formula (\ref{latcurv}).
For the surface on the picture this sum is equal to 
$A = 16+20+20= 56$ times $ a {\pi \over 2}$.
}
\label{fig}
\end{figure}

\begin{figure}
\centerline{\hbox{\psfig{figure=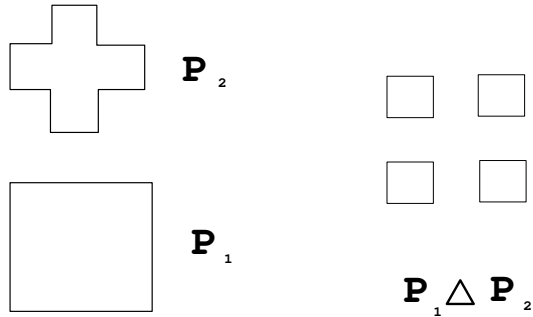,width=7cm}}} \caption[fig3]
{ The transfer martix (\ref{tranmat}) can be 
viewed as describing the propagation of the polygon-loop 
$P_{1}$ at time $\tau$ to another polygon-loop $P_{2}$ at 
the time $\tau +1$. For the surface on the picture we have 
$k(P_1)=4,k(P_2)=12, l(\emptyset \bigtriangleup 
P_1)=12, l(P_1 \bigtriangleup  P_2)=16,
l(P_2 \bigtriangleup\emptyset)=12$. Summing all these 
quantities in accordance with the formula (\ref{commonact}) 
we shall get $A(M)= 4 + 12 +12+16+12 = 56$ times 
$ a {\pi \over 2}$. This, as it should,  coincides with 
the previous result.
 } \label{fig} 
\end{figure}

The functional $k(Q)$ is the total curvature of the polygon-loop $Q$ which is 
equal to the number of corners of the polygon ( the vertices with 
self-intersection are not counted!) and $l(Q)$ is the length of $Q$ 
which is equal to the number of its links.  The  length functional 
$l(Q_{1} \bigtriangleup  Q_{2})$ was defined above (\ref{common}) . The operation
$\bigtriangleup$ maps two polygon-loops $Q_{1}$ and $Q_{2}$
into a  polygon-loop $Q= Q_{1} \bigtriangleup  Q_{2}$ \footnote{Note that the  
operations $\cup$ and $\cap$ do not have this property. 
These operations acting on a polygon-loops can produce link 
configurations 
which do not belong to $\Pi$. The symmetric difference 
of sets $Q_{1} \bigtriangleup  Q_{2}$ is an important concept in 
functional analysis \cite{kolmogorov}.}. The 
length functional $l(Q_{1} \bigtriangleup  Q_{2})$  defines a 
distance between two  polygon-loops $Q_{1}$ and $Q_{2}$.

The eigenvalues of the transfer matrix $K_{\beta}(Q_{1},Q_{2})$ define all 
statistical properties of the system and can be found as a solution of the 
following integral equation in the loop space 
$\Pi$ \cite{transferma}
\be
\sum_{\{Q_{2}\} }K_{\beta}(Q_{1},Q_{2})~\Psi(Q_{2})=
\Lambda(\beta)~\Psi(Q_{1})  \label{inte},
\qee
where $\Psi(Q)$ is a function on loop space. The Hilbert space of 
complex functions $\Psi(Q)$ on $\Pi$ will be denoted as
$H=L^{2}(\Pi)$. The eigenvalues define the partition function (\ref{partfun})
\be
Z^{3d}(\beta)= \Lambda^{N}_{0}+...
+\Lambda^{N}_{\gamma -1} 
 \label{limit},
\qee
and in the thermodynamical limit the free energy is equal to 
\be
-\beta~ f_{3d}(\beta) = \lim_{N \rightarrow \infty} 
\frac{1}{N^3}~ln~ Z^{3d}(\beta)  =  
\lim_{N \rightarrow \infty}\frac{1}{N^2}~ln~ \Lambda_0.
\qee
Finite time propagation amplitude of an initial loop $Q_{i}$ to 
a final loop $Q_{f}$ for the time interval  $t = M/ \beta$ can be 
defined as 
\be
K(Q_{i},Q_{f}) = \Lambda^{-M}_{0} 
\sum_{\{Q_{1},Q_{2},...,Q_{M-1}\}}~ K_{\beta}(Q_{i},Q_{1})\cdots 
K_{\beta}(Q_{M-1},Q_{f}), \label{prop}
\qee
where we have introduced natural normalization to the biggest eigenvalue 
$\Lambda_{0}$ and $M \leq N$. 

{\it Loop Transfer Matrix for $\kappa \neq 0$.}
We would like to generalize the construction of the transfer matrix 
to the case of nonzero self-intersection coupling constant $\kappa$.
As we already mentioned above surfaces on a lattice may have 
self-intersections and we associate the coupling constant $\kappa$ 
with these self-intersection edges of the surface. When we consider 
the sections of the lattice surface $M$ by the planes  $E$
the self-intersection edges will 
appear on a plane as  polygon-loops with self-intersections. When $\kappa=0$ 
we have been instructed to compute the total curvature $k(E)$ of these 
polygon-loops simply ignoring angles 
at the self-intersection points. Now when $\kappa \neq 0$ we have to take 
into account self-intersection vertices of polygon-loops. 

The contribution to the action which comes from simple polygon right angles
is the same as before (\ref{commonact})
\be
\sum_{\{E_z \}} k(Q_{i}) + l(Q_{i} \bigtriangleup  Q_{i+1}) .\label{kzero}
\qee
The contribution coming from self-intersections should be divided into  
contribution coming from vertical and horizontal edges of the surface.
The vertical and horizontal edges of the surface are defined relative to
the "time" planes $E_z$. The contribution from vertical edges is simply 
the number of self-intersection vertices of loops on $E_z$ planes
\be
4 \kappa \cdot \sum_{\{E_z \}} k_{int}(Q_i),\label{knonzerover}
\qee 
where $k_{int}(Q)$ is the number of self-intersection vertices of loop $Q$.

To count contribution from horizontal edges of the surface one should introduce
the orientation of the polygon-loops on every $E_z$ plane. The same orientation is 
ascribed to all polygon-loops on a given plane $E_z$: clock-wise or anti-clock-wise. 
Orientation is automatic if $Q$ is thought as interface 
between spins. Conjugation $\stackrel{\rightarrow}{Q}$ $\rightarrow$ 
$\stackrel{\leftarrow}{Q}$ is equivalent to $Z_2$ spin flip.
So there is $\stackrel{\rightarrow}{Q}$ and $\stackrel{\leftarrow}{Q}$ with 
opposite orientation.

Now let us consider oriented polygon-loops on two planes $E^{i}_{z}$ and 
$E^{i+1}_{z}$. Then one should select that bonds on these polygon-loops which are 
common. Only bonds with the opposite  orientation should be counted. The 
number  of common bonds with opposite  orientation we shall denote as 
$l(Q_{i}  \stackrel{\leftrightarrow}{\cap}  Q_{j})$. The notation $\cap$
is used to denote the common bonds, the symbol 
$\leftrightarrow$ on the top of it to 
denote that they should have opposite orientations. Therefore the contribution 
from horizontal edges can be expressed as   
\be
4 \kappa \cdot  \sum_{\{E_z \}} l(Q_{i}  \stackrel{\leftrightarrow}{\cap}  Q_{j}) 
\label{knonzerohor}
\qee
and the  contribution from all self-intersections as 
\be
4 \kappa \cdot \sum_{\{E_z \}}~ k_{int}(Q_i) +  l(Q_{i}  
\stackrel{\leftrightarrow}{\cap}  Q_{i+1}). 
\label{knonzerovh}
\qee  
The total action (\ref{energyk}) now takes the form:
\be
A(M) = \sum_{\{E_z \}} ~ k(Q_{i}) + l(Q_{i} \bigtriangleup  Q_{i+1}) +  
4\kappa \cdot[~ k_{int}(Q_i) +  l(Q_{i}  \stackrel{\leftrightarrow}{\cap}  Q_{i+1})~] .
\label{totknonzero}
\qee
The partition function (\ref{partfun}) can be now represented in the 
same form (\ref{trace}) where $K_{\beta}(Q_{1},Q_{2})$ is again the 
transfer matrix of  size $\gamma \times \gamma $, defined as 
\bea
K_{\beta}(Q_{1},Q_{2})~~~~ = ~~~~exp \{-\beta ~[k(Q_{1}) + 
2 l(Q_{1} \bigtriangleup  Q_{2}) + k(Q_{2})] \}\cdot~~~~~~~~~~ \nonumber\\
exp \{-4\kappa \beta ~[ k_{int}(Q_1) + 2 l(Q_{1}  
\stackrel{\leftrightarrow}{\cap}  Q_{2}) +  k_{int}(Q_2) ]\},  
\label{nonzerotranmat}
\eea
where $Q_{1}$ and $Q_{2}$ are closed oriented polygon loops on a two-dimensional 
lattice 
$T^{2}$ of size $N \times N$ and $\gamma$ is the total number of polygon-loops 
on a lattice $T^{2}$. Let us now consider the limit 
$$
\kappa \rightarrow \infty,
$$
then from our expression for the transfer matrix (\ref{nonzerotranmat}) one can 
conclude that on every time slice there will be no polygon-loops  with 
self-intersections. This means that they are just nonsingular self-avoiding 
oriented loops. In addition there is strong correlation between  loops 
on consecutive planes: common bonds with opposite  orientation are 
also strongly suppressed. Which means that propagation of oriented loop in time
has small probability to invert its orientation.

\section{Loop Fourier transformation}

In the approximation when we drop the
curvature term in the expression (\ref{tranmat}) for the loop
transfer matrix $K(Q_{1},Q_{2})$
the spectrum of the  matrix has been evaluated exactly
in the articles \cite{transferma,sav2}. 
In this section I shall present an alternative method of derivation 
which is based on generalization of Fourier transformation and allows 
to carry out analogy with quantum mechanics of point particles and to emphasize
generality of the approach. 

I shall consider two different approximations 
to the initial transfer matrix (\ref{tranmat}) 
suggested in  \cite{sav2} and \cite{transferma}.
In the first one  it was suggested to substruct 
the term $2 k(Q_1 \cup Q_2)$  and to consider the transfer matrix of the 
form \cite{sav2} 
\be
\tilde{K}(Q_{1} \bigtriangleup  Q_{2}) = 
exp \{-\beta ~[k(Q_{1}\bigtriangleup  Q_{2}) + 
2 l(Q_{1} \bigtriangleup  Q_{2})]~ \}, \label{firstapp}
\qee 
and in the second one  to consider the transfer matrix 
\cite{transferma}
\be
K(Q_{1} \bigtriangleup  Q_{2}) = 
exp \{-2 \beta ~l(Q_{1} \bigtriangleup  Q_{2})~ \}. \label{secondapp}
\qee 
In both cases the transfer matrix depends only on the symmetric difference  
$Q_{1} \bigtriangleup  Q_{2}$ and the integral equation  (\ref{inte}) take 
the form
\be
\sum_{\{Q_{2}\} }K_{\beta}(Q_{1} \bigtriangleup  Q_{2})~\Psi(Q_{2})=
\Lambda(\beta)~\Psi(Q_{1})  \label{differinte}.
\qee
The fact that transfer matrix now depends only on the symmetric difference 
is of paramount importance in solving these models. 
The reason is that we can make a
change of variable, from $Q_2$ to $Q$ at fixed $Q_{1}$ \cite{transferma} 
\be
Q = Q_1  \bigtriangleup  Q_{2} \label{newva}
\qee
and  see that equation transforms to the form
\be
\sum_{\{Q\} }K_{\beta}(Q)~\Psi(Q \bigtriangleup  Q_{1})=
\Lambda(\beta)~\Psi(Q_{1}).  \label{loopeq}
\qee
The integral equation will be solved  if we can 
find the set  of loop functions $\Psi(Q)$ which satisfy the 
following functional equation on  loop space $\Pi$
\be
\Psi(Q \bigtriangleup  Q_{1}) = \Psi(Q) \cdot \Psi(Q_{1}). \label{loopfun}
\qee
For these functions the equation (\ref{loopeq}) reduces 
to 
\be
\sum_{\{Q\} }K_{\beta}(Q)~\Psi(Q) =  \Lambda(\beta) \label{egenval}
\qee
and we can find all eigenvalues. The constant function $\Psi =1$ is the solution of the 
functional equation (\ref{loopfun}) and due to Frobenius-Perron theorem 
corresponds to the largest and non-degenerate  eigenvalue $\Lambda_\emptyset$ 
of the symmetric transfer matrix $K(Q_1 \bigtriangleup  Q_{2})$.

The functional equation (\ref{loopfun}) is analogous to the equation 
$\psi(x+y)=\psi(x)\cdot\psi(y)$ appearing in 
ordinary quantum mechanics, which allows to find all eigenvectors 
and eigenvalues of the integral equation
\be
\int K_{\beta}(x-y) \psi(y) dx = \lambda(\beta) \psi(y),~~~~~~~~
K_{\beta}(x-y)= exp(-\beta(x-y)^2) 
\qee
with the well known solutions
$$
\psi(x) = e^{ipx},~~~~~~~~~~~~\lambda_{p}(x) = e^{-p^2/4\beta}.
$$
Our idea is to find all solutions of the functional equation (\ref{loopfun})
and the analog of the plane wave solution in loop space.
From functional equation (\ref{loopfun}) it follows that if $Q = Q_1$
then
\be
\Psi(\emptyset) = \Psi^{2}(Q)
\qee
and if $Q = \emptyset$ then $\Psi(\emptyset)$ is equal to zero or to one,
therefore  nontrivial solution is
\be
\Psi^{2}(Q) = 1~~~~~  \Rightarrow ~~~~~\Psi(Q) = \pm 1.
\qee
This suggests the following form of the eigenfunctions
\be
\Psi_{P}(Q) = e^{i\pi s(P \cap Q) } \label{loopqm}
\qee
which are numbered by the loop momentum $P$ and take values $\pm 1$.
The $s(Q)$ is the area of the region with the boundary loop $Q$. 
These solutions are in pure analogy with plane
waves in quantum mechanics . To check that they are indeed solutions 
of the functional equation (\ref{loopfun}) we have to use the set identity
\be
(Q_1 \bigtriangleup Q_2) \cap P = (Q_1 \cap P) \bigtriangleup (Q_2 \cap P)
\qee
which is easy to verify and the relation $s(Q_{i})+s(Q_{i+1})-2 \cdot s(Q_{i} \cap Q_{i+1} ) 
= s(Q_{i} \bigtriangleup  Q_{i+1})$. 
To carry out the analogy with plane wave solution 
of quantum mechanics deeper one should 
interpret the loop $P$ as a {\it loop momentum} conjugate to a 
loop coordinate variable $Q$. To prove orthogonality of these loop wave 
functions let us compute the product
\be
\Psi_{P_{1}}\cdot \Psi_{P_{2}} = 
\sum_{\{Q\}} \Psi_{P_{1}}(Q)~ \Psi_{P_{2}}(Q)
=\sum_{\{Q\}} e^{i\pi s(P_1 \cap Q)  +  i\pi s(P_2 \cap Q) }
=\sum_{\{Q\}} e^{i\pi s((P_1 \bigtriangleup P_2) \cap Q)  }
\label{orthogonality}
\qee
where we again used the above set identity and the relation (\ref{common})
for $s(Q)$.  The last sum is 
equal to zero if $P_1 \bigtriangleup P_2 \neq \emptyset$ and
is equal to the number of loops $\gamma$ on the lattice if 
$P_1 \bigtriangleup P_2 = \emptyset$, thus
\be
\sum_{\{Q\}} e^{i\pi s((P_1 \bigtriangleup P_2) \cap Q)} = 
2^{N^2} \delta (P_1 , P_2).
\qee
The last statement follows from the fact that the product 
between the constant function $\Psi_{\emptyset} =1$, corresponding to  
the largest and non-degenerate  eigenvalue $\Lambda_\emptyset$, and
all others $exp(i\pi s(P \cap Q))$ is equal to zero if
$P$ is nonempty 
\be
\sum_{\{Q\}}1 \cdot e^{i\pi s((P \cap Q)  } = 0,~~~~if~~~~~P \neq \emptyset.
\qee
The number of functions in this orthogonal set is equal therefore to 
the number of closed loops $P$, that is $\gamma = 2^{N^2}$.

Having in hand all eigenfuctions we can find all eigenvalues from (\ref{egenval})
for  transfer matrices (\ref{firstapp}) and (\ref{secondapp})
\be
\tilde{\Lambda}_{P} = \sum_{\{Q\} } 
e^{-i\pi s(P \cap Q) -  \beta[k(Q) + 2 l(Q)]}~~~~~~~~~~\Lambda_{P} 
= \sum_{\{Q\} } e^{-i\pi s(P \cap Q) - 2 \beta l(Q) }.
\qee
Because the expression $\sum_{\{Q\} } e^{ - 2 \beta l(Q)} $ is the 
partition function of the 2d Ising model \cite{kramers,baxter} and 
$\sum_{\{Q\} } e^{ - \beta [2 l(Q)  + k(Q)] } $ is the 
partition function $\tilde{Z}^{2d}$ of 
the two-dimensional model solved in \cite{sav2} we  see that 
the largest eigenvalue $\Lambda_{\emptyset}$ 
is exactly equal to a corresponding partition function
\be
\tilde{\Lambda}_{\emptyset} = \tilde{Z}^{2d}~~~~~~~~~~\Lambda_{\emptyset} 
= Z^{2d-Ising}.
\qee
For the ratio of eigenvalues we shall have
\footnote{In what follows we shall consider the transfer 
matrix (\ref{secondapp}). 
Analogous formulas are valid for the transfer 
matrix (\ref{firstapp}) with curvature term $k(Q)$.}
\be
{\Lambda_{P} \over \Lambda_{\emptyset} } ~= ~ \sum_{\{Q\} } 
e^{-i\pi s(P \cap Q) - 2 \beta l(Q) }/Z^{2d-Ising} ~\equiv ~ <e^{-i\pi s(P \cap Q)}>_Q
\qee
which is identically equal to the spin correlation functions of 
the 2d Ising model  
\be
{\Lambda_{P_{\vec{r_1},...\vec{r_n}}} 
\over \Lambda_{\emptyset} }  ~ = ~ <\sigma_{\vec{r_1}}\cdot 
\cdot \cdot \sigma_{\vec{r_n}}>_{2dIsing}.
\qee
In this formula the loop momentum $P$ is defined by the 
vectors $\vec{r_1},...\vec{r_n}$.  In particular for the 
smallest (one-box) loop momentum $P= \Box$ we shall have
\be
{\Lambda_{\Box} \over \Lambda_{\emptyset} } ~=~ <\sigma_{\vec{r}}>~=~\mu(\beta)
\qee
which coincides with the magnetization of 2d Ising model. Because it does
not depend on position of the loop on the lattice, this eigenvalue 
is $N^2$ degenerate. For the two-box loop momentum $P =~\sqsubset\sqsupset$ 
consisting of two neighboring boxes we shall have 
\be
{\Lambda_{\sqsubset\sqsupset} \over 
\Lambda_{\emptyset} } ~=~ <\sigma_{\vec{r}} \sigma_{\vec{r + 1}} >~=~-u(\beta)
\qee
which is $2N^2$ degenerate and after summation over all $2N^2$ bonds of the 
lattice we shall get the internal energy of the 2d Ising model.

We can also use loop wave functions to define generalized loop-Fourier
transformation as 
\be
\Psi(P) = {1 \over 2^{N^2} } \sum_{\{Q\}}  e^{i\pi s(P \cap Q)} \Psi(Q)
\qee
and then compute the loop Fourier transformation of the transfer 
matrices (\ref{firstapp}) and (\ref{secondapp}) as:
\bea
K(P_1 , P_2) = \sum_{\{Q_1 , Q_2\}} K(  Q_{1} \bigtriangleup  Q_{2})
e^{i\pi s(P_1 \cap Q_1)  -  i\pi s(P_2 \cap Q_2) } 
= \Lambda_{P_1} \sum_{\{Q_2\}}  
e^{i\pi s(P_1 \cap Q_2)  -  i\pi s(P_2 \cap Q_2) }\nonumber\\
= \Lambda_{P_1} \delta (P_1 , P_2)
.~~~~~~~~~~~~~~~~~~~~~~~~~~~~~~~~~~~~~~~~~~~~~~ \label{diagonal}
\eea
The finite time propagation amplitude of an initial loop $P_{i}$ to 
the final loop $P_{f}$ for the time interval  $t = M/ \beta$ (\ref{prop}) we 
have  
\bea
K(P_{i},P_{f}) =({\Lambda_{P_i} \over 
\Lambda_{\emptyset}})^{M}~~\delta (P_i , P_f)
=e^{M ln ({ \Lambda_{P_i} \over \Lambda_{\emptyset} } ) }~~ \delta (P_i , P_f)=
e^{-M H(P_i)}~~ \delta (P_i , P_f),
 \label{propag}
\eea
therefore for the loop Hamiltonian we shall get
\be
H(P)= -ln ({ \Lambda_{P} \over \Lambda_{\emptyset}}) 
\qee
where $H(P)$ is the energy of the coherent state with loop momentum $P$. 

{\it Spin system which corresponds to the matrix 
$K(Q_{i} \bigtriangleup  Q_{i+1} )$ (\ref{secondapp}). }
The natural question which arises here is how nontrivial is the 3d system 
which has been solved in this way? To make the answer as clear as 
possible we should go back and express the system 
which has been solved again in terms of  spins 
on a 3d lattice and to see what is left out of 
original interactions.  

\begin{figure}
\centerline{\hbox{\psfig{figure=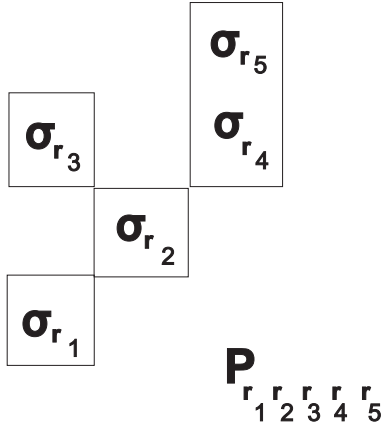,width=5cm}}}
\caption[fig8]{The loop momentum $P_{\vec{r_1},...\vec{r_n}}$ is defined by the 
vectors $\vec{r_1},...,\vec{r_n}$ and the corresponding eigenvalue 
$\Lambda_{P_{\vec{r_1},...\vec{r_n} }}$ is equal to 
the correlation function of the $n$ spins ~~$~~<\sigma_{\vec{r_1}}\cdot \cdot 
\cdot \sigma_{\vec{r_n}}>_{2d}$.}
\label{fig8}
\end{figure}

We ignore the curvature  terms $k(Q_i)$ in 
the transfer matrix, this means that we ignore the contribution coming 
from the edges of the surface which are along the time direction, 
or $z$ direction.
The contribution coming  from the edges which are parallel to the $x$ and $y$
direction are left and their contribution is equal to (\ref{common})
\be
\sum_{\{ E_{z} \}} l(Q_{i} \bigtriangleup  Q_{i+1}).
\qee  
In terms of spin variables the original action can be 
expressed as a sum of energies coming from all edges of the surface (see formulas 
(5a) and (7) in \cite{weg})  
\be
H = \sum_{over~all~edges} H_{edge}, \label{ham}
\qee
where
\be
H_{edge} = U_{1}U_{-1} + U_{1}U_{-1} = 
\sigma_{1}\sigma_{2}\sigma_{-1}\sigma_{-2} \label{fourspin}.
\qee
is a four-spin interaction term with spins distributed around the edge.
Because we ignore the edges in the $z$ direction, we have to sum 
in (\ref{ham}) only over $A_{edges}$  in  $x$ and $y$ direction. This 
corresponds to the summation over four-spin interaction terms 
(\ref{fourspin}) which lie only on the planes $E_x$ and $E_y$. These 
planes are normal to the $x$ and $y$ direction. Therefore 
our approximation corresponds to the 3d lattice spin system  
in which  four-spin interactions take place only on vertical 
planes $E_x$ and $E_y$
\be
H_{Q_{1} \bigtriangleup  Q_{2}} = \sum_{E_{x},E_{y}}\sigma\sigma\sigma\sigma. 
\qee
It is obvious that this spin system can not be factored 
into non interacting two-dimensional subsystems. We 
can ask what is the effective interaction between 
spins which lie on a given two-dimensional $E_z$ plane?
Let us label spins which are on that plane as $\sigma$,
spins which are on the previous plane as $\lambda$, 
and spins which are on the next plane as $\mu$. The 
part of the Hamiltonian which contains only $\sigma$ spins 
can be written in the form:
\be
\sum_{i} \sigma_{i}\sigma_{i+1}\lambda_{i}\lambda_{i+1} +
\sigma_{i}\sigma_{i+1}\mu_{i}\mu_{i+1} = 
\sum_{i} (\lambda_{i}\lambda_{i+1} + \mu_{i}\mu_{i+1})
\sigma_{i}\sigma_{i+1} = \sum_{i} J_{eff}\sigma_{i}\sigma_{i+1},
\qee   
where the effective coupling $J_{eff}$ depends on 
spins on the neighboring planes. It 
takes the values $-2,0,2$ with probabilities $1/6,4/6,1/6$ 
simply because 
$$
J_{eff} = \lambda_{i}\lambda_{i+1} + \mu_{i}\mu_{i+1}.
$$ 
For the observer who lives on the plane 
$E_z$  spin interactions  
look like 2d Ising model with randomly distributed 
coupling constant $J_{eff}$.

{\it 3d Ising Transfer Matrix.}
Similar expression for the transfer matrix has been derived in \cite{sav2}
for the 3d Ising model
\be
K^{3d-Ising}_{\beta}(P_{1},P_{2}) = 
exp \{-\beta ~[l(P_{1}) + 
2 s(P_{1} \bigtriangleup  P_{2}) + l(P_{2})]~ \},  \label{3dItranmat}
\qee
where $s(P)$ is the area functional. It is instructive to 
compare 3d-gonihedric and 3d Ising systems. One can see simple 
hierarchical structure of the corresponding transfer 
matrices  (\ref{tranmat}) , (\ref{3dItranmat}). If one 
orders the  functionals in the increasing geometrical complexity
\be
curvature~k(P),~~~~~length~l(P),~~~~~~~~area~s(P), \label{hierarchy}
\qee
then it becomes clear that gonihedric transfer matrix depends on the 
first two functionals: curvature and length, while the 3d Ising 
model depends on the last two functionals:  length and area.
Thus if one puts the models in order of increasing complexity 
one should consider gonihedric model and then Ising model \cite{sav2}. 

Let us also consider similar to (\ref{firstapp}), (\ref{secondapp}) 
approximations to the 3d Ising transfer matrix, they are:
\be 
\tilde{K}(Q_{1} \bigtriangleup  Q_{2}) = 
exp \{-\beta ~[l(Q_{1}\bigtriangleup  Q_{2}) + 
2 s(Q_{1} \bigtriangleup  Q_{2})]~ \},~~~~
K(Q_{1} \bigtriangleup  Q_{2}) = 
exp \{-\beta ~2 s(Q_{1} \bigtriangleup  Q_{2})~ \},
\label{3Dfirstapp}
\qee 
Both approximations can be solved by using the set of eigenfunctions 
(\ref{loopqm})
\be
\tilde{\Lambda}_{P} = \sum_{\{Q\} } 
e^{-i\pi s(P \cap Q) -  \beta[l(Q) + 2 s(Q)]}~~~~~~~~~~\Lambda_{P} 
= \sum_{\{Q\} } e^{-i\pi s(P \cap Q) - 2 \beta s(Q) }.
\qee
The first relation expresses the eigenvalues through the 
2d Ising model in background magnetic field, the model 
which has not been solved yet exactly \cite{polyakov}, and  in
the second case through free spins in background magnetic field.

\section{Matrices depending only on symmetric difference}

It is obvious that the main reason why we have been able to 
reduce solution of three-dimensional statistical system to a 
solution of two-dimensional  system is special
dependence of the transfer matrix argument from symmetric
difference $Q_{1} \bigtriangleup  Q_{2}$.

We are aimed to formulate the following general result: 
all transfer matrices 
of  three-dimensional statistical systems 
which depend only on symmetric 
difference  $Q_{1} \bigtriangleup  Q_{2}$ 
\be
Z^{3d}(\beta) = Tr K^{N}= e^{-\beta f_{3d}(\beta) N^3}
,~~~~~~~K(Q_{1} \bigtriangleup  Q_{2})= 
e^{-\beta \Omega(Q_{1} \bigtriangleup  Q_{2})}
\qee
where $\Omega$ is arbitrary energy functional,
can be diagonalized by using orthogonal set of functions 
$\Psi_{P}(Q)= e^{i\pi s(P \cap Q) }.$
The eigenvalues can be found exactly in the same way as we 
described above and are equal to the following averages 
\be
\Lambda_{P}= \sum_{\{Q\} } e^{-i\pi s(P \cap Q)} 
K(Q)~=~\sum_{\{Q\} } e^{-i\pi s(P \cap Q)  -\beta \Omega(Q)  }. \label{och}
\qee
{\it The eigenvalues $\Lambda_{P}$ of three-dimensional system 
are exactly equal to the correlation functions 
of the two-dimensional system with the partition function}
\be
Z^{2d}(\beta) =\sum_{ \{Q\} } e^{ -\beta \Omega(Q)}= 
e^{-\beta f_{2d}(\beta)\cdot N^2}
=~\lambda^{N}_{0} +...+ \lambda^{N}_{2^{N}-1}, \label{2d}
\qee 
where $f_{2d}(\beta)$ is the free energy of the 2d model and
$\lambda_{i}$ are the eigenvalues of the corresponding
transfer matrix. 
In particular from (\ref{och}) we see that the largest 
eigenvalue $\Lambda_{\emptyset}$ 
is equal to the partition function of $2d$ system 
\be
\Lambda_{\emptyset} = \sum_{\{Q\} } 
e^{-\beta \Omega(Q)  }~ \equiv ~Z^{2d}(\beta) ~=~\lambda^{N}_{0} 
+...+ \lambda^{N}_{2^{N}-1},  
\qee
therefore we can express free energy of 3d system in terms of 
largest eigenvalue $\Lambda_{\emptyset}$ and then through the
partition function of the corresponding two-dimensional 
system
\bea
-\beta~ f_{3d}(\beta) = \lim_{N \rightarrow \infty} 
\frac{1}{N^3}~ln~ Z^{3d}(\beta)  = \lim_{N \rightarrow \infty} 
\frac{1}{N^3}~ln \{ \Lambda^{N}_{\emptyset} +...\}\nonumber\\
=\lim_{N \rightarrow \infty} \frac{1}{N^3}~ln \{
[ \lambda^{N}_{0} +...+ \lambda^{N}_{2^{N}-1}]^N + ...\} =
-\beta~ f_{2d}(\beta).
\eea
Thus we have amazing equality 
$$
f_{3d}(\beta) = f_{2d}(\beta)
$$
which proves identical critical behaviour of 2d and 3d systems! 
Because 
$$
\sum_{\{Q\} } e^{-i\pi s(P \cap Q)  -\beta \Omega(Q)  }/Z^{2d}(\beta) ~=~ <
 e^{-i\pi s(P \cap Q)}>_{Q} = ~ = ~ <\sigma_{\vec{r_1}}\cdot 
\cdot \cdot \sigma_{\vec{r_n}}>_{2d},
$$
from (\ref{och}) it follows that the ration of eigenvalues is
equal to:
\be
{\Lambda_{P_{\vec{r_1},...\vec{r_n}}} 
\over \Lambda_{\emptyset} }  ~ = ~ <\sigma_{\vec{r_1}}\cdot 
\cdot \cdot \sigma_{\vec{r_n}}>_{2d}.
\qee
In this formula the loop momentum $P$ is defined by the 
vectors $\vec{r_1},...\vec{r_n}$. 
It is clear that the correlation functions
of the three-dimensional system are more complicated 
conglomerates.  

It is easy to understand that the result is even more general:
any $d-dimensional$ system which has transfer matrix depending
only on symmetric difference $Q_{1} \bigtriangleup  Q_{2}$ can be 
reduced to a corresponding $(d-1)-dimensional$ system and its
eigenvalues are simply the correlation functions of the 
lower dimensional system. Below  we shall consider an important
example of such dimensional reduction in four-dimensions.

{\it Transfer Matrix for Membranes}
In \cite{sav2} the authors have constructed the  transfer 
matrix in four-dimensions which describes a propagation 
of two-dimensional membrane. Below we 
shall see that this system also can be  solved in terms of 
three-dimensional system considered in previous 
sections.

Transfer matrix which describes the propagation of two-dimensional 
membrane $M$ has the following form \cite{sav2}
\be
K_{\beta}(M_{1},M_{2})~~~~ = ~~~~exp \{-\beta ~[\chi(M_{1}) + 
2 A(M_{1} \bigtriangleup  M_{2}) + \chi(M_{2})] ~\},  
\label{4dtranmat}
\qee
where $\chi(M)$ defines the module of the Euler character 
of the surface $M$
\be
\chi(M) = \sum_{<i>} \vert 2\pi - \omega_{i} \vert ,
\qee 
summation is over all vertices of triangulated 
surface $M$ and $A(M)$ is the gonihedric functional 
which we defined earlier. If we introduce the set of 
orthogonal functions 
\be
\Psi_{P}(M) = e^{i\pi v(P \cap M)}, \label{3dfunc}
\qee
where $v(M)$ is the volume of the region whose boundary is the 
surface $M$, then transfer matrices which depend only on
symmetric difference of surfaces $M_1$ and $M_2$
\be
\tilde{K}(M_{1},M_{2})~ = ~exp \{-\beta ~[\chi(M_{1} \bigtriangleup M_{2}) + 
2 A(M_{1} \bigtriangleup  M_{2}) ] \},~~~~~
K(M_{1},M_{2}) ~ = ~exp \{-2\beta ~
A(M_{1} \bigtriangleup  M_{2}) \},
\qee
can be diagonalized by using the above  set of functions (\ref{3dfunc})
\be
\tilde{\Lambda}_{P} = \sum_{\{M\} } e^{-i\pi v(P \cap M)  
-\beta [\chi(M) +  2A(M)] },~~~~~\Lambda_{P} = 
\sum_{\{M\} } e^{-i\pi v(P \cap M)   -2 \beta A(M) }. \label{dimreduc}
\qee
For $P = \emptyset$ we have 
\be
\Lambda_{\emptyset} = \sum_{\{M\} } e^{ -2 \beta A(M) }
\qee
and it coincides with the partition function of the gonihedric 
system (\ref{partfan}).
The free energy of this four-dimensional model is equal therefore to
\be
f_{4d}(\beta) = f_{3d}(\beta) = f_{2d-Ising}(\beta) 
\qee
and the eigenvalues (\ref{dimreduc}) are the correlation function of 
the 3d gonihedric system 
\be
(   {\Lambda_{P} \over \Lambda_{\emptyset} } )^{4d}~= ~ (Correlation~functions)^{3d}.
\qee
Thus the eigenvalues of the transfer matrix in four-dimensions  can be described 
by the correlation functions of the system of one less dimension.

\section{Discussion}

We have seen that transfer matrix approach allows to solve large 
class of statistical systems when transfer matrix depends on 
symmetric difference of propagating loops, membranes or p-branes. The 
eigenvalues are equal to the correlation functions of the corresponding statistical
system, which has smaller dimension than the original one. This dimensional 
reduction expresses the hierarchical structure in which more 
complicated high-dimensional systems are a superposition
of lower-dimensional ones \cite{sav2}. This structure is especially visible   
when one uses generalization of Fourier transformation in loop
space or in p-brane spaces.

In conclusion I would like to acknowledge R.Kirschner, B.Geyer and 
W.Janke for discussions and kind hospitality at Leipzig University.
This work was supported in part by the EEC Grant no. HPMF-CT-1999-00162.

\vfill
\end{document}